\documentclass[fleqn,usenatbib]{mnras}

\usepackage{newtxtext,newtxmath}

\usepackage[T1]{fontenc}

\DeclareRobustCommand{\VAN}[3]{#2}
\let\VANthebibliography\thebibliography
\def\thebibliography{\DeclareRobustCommand{\VAN}[3]{##3}\VANthebibliography}

\usepackage{graphicx}	
\usepackage{amsmath}	
\usepackage{amssymb}	
\usepackage{pdflscape}	
\usepackage{soul}

\usepackage{xcolor} 

\title[ASAS Census of Twins]{ASAS Census of Twins}

\author[Bak{\i}\c{s} et al.]{
Volkan Bak{\i}\c{s}$^{1}$\thanks{E-mail: volkanbakis@akdeniz.edu.tr},
Zeki Eker$^{1}$,
Oguzhan Sar{\i}$^{2}$,
G\"{o}khan Y\"{u}cel$^{1}$
and Eda Sonba\c{s}$^{3}$
\\
$^{1}$Akdeniz University, Department of Space Sciences and Technologies, Dumlupinar Blv., Kamp\"{u}s, 07058, Antalya, TR\\
$^{2}$Milli Savunma University, Kara Astsubay Meslek Y\"{u}ksekokulu, Alt{\i}eyl\"{u}l, Bal{\i}kesir, TR \\
$^{3}$University of Ad{\i}yaman, Department of Physics, 02040, Ad{\i}yaman, TR
}

\date{Accepted 2020 June 3. Received 2020 May 29; in original form 2020 February 20}

\pubyear{2020}

\begin{document}
\label{firstpage}
\pagerange{\pageref{firstpage}--\pageref{lastpage}}
\maketitle

\begin{abstract}
Twin binaries were identified among the eclipsing binaries with  $\delta$>--30$^\circ$ listed in the ASAS catalog. In addition to the known twin binaries in the literature, 68 new systems have been identified, photometric and spectroscopic observations were done. Color, spectral type, temperature, ratio of radii and masses of the components have been derived and presented. Including 12 twin binary systems that exist in both ASAS and the catalogue of absolute parameters of detached eclipsing binary stars, a total of 80 twin detached binary systems have been statistically studied. A comparison of the spectral type distribution of the twins with those of detached eclipsing binary stars in the ASAS database showed that the spectral type distribution of twins is similar to that of detached systems. This result has been interpreted as there is no special formation mechanism for twins than for normal detached binaries. As a result of our case study for HD~154010, a twin binary, we presented the precise physical parameters of the system.
\end{abstract}




\section{Introduction}

The first statistical study of binary stars with known mass ratio was conducted by \cite{LucyRicco1979} when the binary stars were started to be explained with the scenarios including numerical solutions of fission and fragmentation. They studied the mass ratio-probability density (q-N) of late B-type, Solar-type and RS CVn-type binaries, by taking into account factors such as radial velocity (RV) measurement errors, line identification errors, selection effect, etc., which may create a systematic error in determining the mass ratio (see Figures 4,5 and 6 in \cite{LucyRicco1979}). In the same study, they found a thin peak around the mass ratio q $\approx$ 0.97 for all three binary groups. The data set they used is free from stellar evolution and selection effects, therefore the thin peak in the q-N graph can only be explained by the star formation mechanism. \cite{LucyRicco1979} also evaluated two different star formation mechanisms and note that fission can form binary stars with a mass ratio of q $\approx$ 0.2, and that binary stars with q $\approx$ 1.0 can only be formed by fragmentation. According to this bimodal binary star formation scenario, the formation of equal mass binary stars is explained by a split that occurs during the final stages of cloud collapse during the formation. However, modern theoretical simulations \citep{Bate1998} do not confirm such a division. Another formation scenario, accretion, is based on the general acceptance that pre-main sequence stars continue to interact with the surrounding matter as they move toward the main sequence, and the orbit decay needs to shrink during this interaction. \cite{Bate1998} showed that the mass ratio in the formation of close-binary stars has a strong tendency to add to $q=1$ by theoretical models. However, in order for this model to work well, the two stars must have gathered enough mass during the formation and the components should be close enough to each other.

Nevertheless, \cite{Tokovinin2000} looked at the role of multiple systems in the formation scenario, since most twin binary stars are known to have a distant third body, indicating that all other twin binaries may have at least one distant component. It is known that, in the early stages of the formation, dynamic interactions in multiple systems can form binaries with mass ratio close to one. However, the fact that twin binary stars appear frequently in multiple systems does not indicate a direct relationship between twin pairs and multiple systems, but it is a common factor in the formation of multiple systems with twin binaries \citep{Tokovinin2000}.

\cite{Halbwachs2003} first mentioned spectral type dependence of q-N diagrams of spectroscopic binary stars. They examined the mass ratio distribution of spectroscopic binaries of F-G and K spectral types and concluded that these three spectral types showed a similar distribution except for small deviations. More recent studies on the mass ratio distributions of binary stars (e.g. \cite{Lucy2006}; \cite{SimonObbie2009}) show that the proportion of twins with a mass ratio of $q = 0.98-1.00$ is around 3 percent and that F, G and K spectral types are dominant. \cite{SimonObbie2009} include 1 O, 1 B, 3 A, 16 F, 11 G and 3 K-type stars in their own twin binary star list. This statistic led them to suggest that small mass twins are in binary star populations where the aggregation process in star formation is slower. In addition, the presence of early-spectral type twins suggests that binary stars with masses greater than 1.6 solar mass may have the same formation mechanism as binary stars of smaller mass stars \citep{ZinneckerYorke2007} but different mechanisms may be important \citep{SimonObbie2009}. For this reason, it is clear that the determination of early-spectral types among twin binary stars will contribute in general to the understanding of star formation and in particular to the understanding of formation of two stars with the same characteristics and testing of current star formation models.

Although the studies focused on the q-N distribution of binary stars have agreed on a bimodal scenario for star formation, the most recent studies (e.g. \cite{Moe2017}, \cite{ElBadry2019} and \cite{Tokovinin2020}) showed that the most of the close binary stars and triple systems formed by disc fragmentation followed by accretion-driven inward migration.

We have noticed that apart from limited examinations of some individual stars, no observational studies have been conducted on twin binary stars in a systematic way. Responding to this requirement is the primary motivation of this study. Therefore, the aim of this paper is to see the spectral type distribution of twin binaries and to investigate if the twin binary formation mechanism differs from that of other close binaries by making systematic observations of twin eclipsing detached binary stars.

The All Sky Automated Survey (ASAS) \citep{Pojmanski1997} catalogue has been selected as the most suitable database for our purpose (\S2.1). After examining the photometric data in the database and identifying the twin candidates (\S2.2), systematic spectral and photometric observations of these candidates were performed (\S2.3). Spectral and photometric observational data were analyzed and spectral type and light curve (LC) parameters were obtained (\S2.4-5). Thus, a final list of twins was created and analyzed statistically (\S3-4). In  \S5, a full spectral and photometric solutions of the A-type twin binary system HD~154010 in our list has been made. Finally, a discussion on our findings and conclusions are presented in \S6.

\section{Methodology}

Young detached binaries which have not yet gone through mass transfer must be well inside their Roche lobes. Since mass exchange between the components during detached phases is not possible, the components must retain their initial masses. If mass loss is negligible during main-sequence lifetime, the main-sequence mass-luminosity relation (MLR) must be in play. The most recent MLR for main-sequence stars in the mass range 0.179-31\,M$_\odot$ is obtained by \cite{Eker2018}. In their study, they have determined different exponential values ($L\,\propto\,M^\beta$) for different mass ranges (see Table~\ref{tab:exponents}). The stellar luminosity is estimated from Eq.~\ref{eq:luminosity}.

\begin{table}
\renewcommand{\thetable}{\arabic{table}}
\centering
\caption{MLR exponentials for different mass ranges.}
\begin{tabular}{cc}
\hline \hline
Mass Interval & Exponent \\
              & ($\beta$) \\
\hline
 0.179$<$\,$M/M_\odot$\,$\leqslant$0.45   &  2.028  \\
 0.45$<$\,$M/M_\odot$\,$\leqslant$0.72   & 4.572   \\
 0.72$<$\,$M/M_\odot$\,$\leqslant$1.05   & 5.743   \\
1.05$<$\,$M/M_\odot$\,$\leqslant$2.40 & 4.329   \\
2.40$<$\,$M/M_\odot$\,$\leqslant$7.0 & 3.967   \\
7$<$\,$M/M_\odot$\,$\leqslant$31 & 2.865   \\
\hline \label{tab:exponents}
\end{tabular}
\end{table}

\begin{equation}
\label{eq:luminosity}
L = 4 \pi R^2 \sigma T_{eff}^{4}
\end{equation}

where $R$ is the stellar radius and $T_{eff}$ is the effective temperature. Because both star formed in the same chemical environment it is most likely that they have the same metallicity and we therefore ignored the effect of the metallicity. With some arithmetic, the relationship between mass ratio and relative parameters of the components of binary stars can be revealed (Eq.~\ref{eq:massratio}) as,

\begin{equation}
\label{eq:massratio}
q (M_1/M_2) = [(\frac{r_1}{r_2})^2 (\frac{T_{eff_1}}{T_{eff_2}})^4]^\frac{1}{\beta}
\end{equation}

where $r=\frac{R}{a}$ is the fractional radius and $a$ is the semi-major axis.

Does $q\sim$1 found by Eq.~\ref{eq:massratio} mean both components have similar masses? According to stellar evolution, massive stars evolve faster. Main-sequence evolution of massive (M$\,>1M_\odot$) stars occur in a way that increase of luminosity cannot compensate the increase of radius, so the surface temperature decreases except the final stages of main-sequence evolution where central convection zone starts to shrink and disappear by the depletion of hydrogen in the core \citep{Clayton1968}. For solar and less massive stars, however, surface temperature rises initially but drops a little later while both luminosity and radius are increasing. For twin binaries (q$\sim$1) it is expected that both temperature and radius ratio must stay close to one, thus, the right hand side of Eq.~\ref{eq:massratio} as well. Otherwise, for massive stars, if q<{}<1, increase of radius ratio could be compensated by decrease of temperature ratio may result luminosity ratio close to one, which may be confused by components of equal luminosities imitating a twin binary. In order to avoid such cases, one must assure both radius and temperature ratios in Eq.~\ref{eq:massratio} must be close to one, that is, luminosity ratio, which is one or close to one, does not always mean the system is twin binary. This rule is valid for all systems not experienced mass transfer yet.

From this point of view, mass ratio of the system could be estimated from the luminosity ratio of the components, which is determined from observable radius and temperatures of component stars.

\subsection{The Database}

Photometric data in the ASAS (The All Sky Automated Survey) \citep{Pojmanski1997} Catalogue of Variable Stars\footnote{http://www.astrouw.edu.pl/asas/?page=acvs} were used to identify the twin binary candidates. The main reason for choosing this database is that it contains the longest photometric monitoring data set with a reasonable precision (0.05 mag in many cases) of all sky so far. The project, which started with ASAS-1 in 1997, has continuously renewed the observational equipments\footnote{http://www.astrouw.edu.pl/asas/?page=history} and has continued until today to collect photometric data of objects between 8$<V<$14 magnitude in the sky of where it has been located. The new version of the survey (ASAS-4) is located in the Northern hemisphere but its database is not public yet\footnote{http://www.astrouw.edu.pl/asas/?page=catalogues}. Therefore, the limitations of available database are that the brightness limit and targets' location restricted to the Southern hemisphere which makes it possible to observe database objects up to +28 degrees declination. Assuming that the twin binary stars are homogeneously distributed in sky and space, choosing certain part of the sky and certain brightness limits will not introduce selection effect related to sky and brightness distributions of the twin binaries.

\subsection{Determination of Twin Candidates}

\begin{figure}
\begin{center}
\begin{tabular}{c}
\includegraphics[width=\columnwidth]{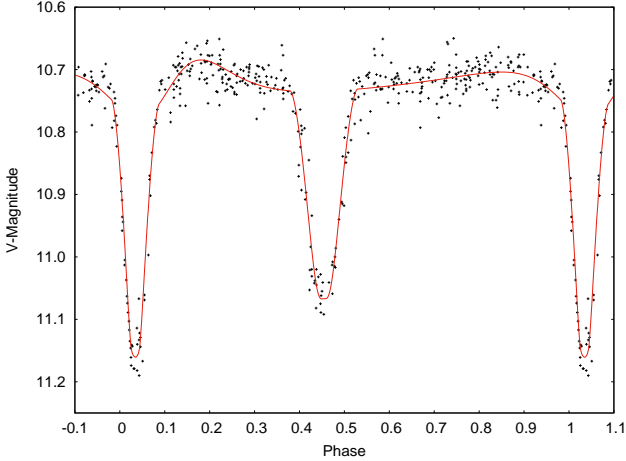}
\end{tabular}
\end{center}
\caption{LC model for the eccentric binary ASAS 073218--1436.5. \label{fig:asas_lee}}
\end{figure}

Before starting to form a candidate list of twins, we recall the study by \cite{Lee2015}, who scanned the ASAS database with specific computer codes and analyzed the light curves by using the approach of \cite{Devor2005}, determined 1558 detached binary systems and derived the mass ratio and component masses. It may appear selecting systems with mass ratio close to one is a straightforward and easy. However, reanalysis of the light curves in \cite{Lee2015} suggests some inconsistencies. For example, reanalysis of the eccentric binary ASAS 073218--1436.5 (see Fig.~\ref{fig:asas_lee}) revealed the temperature ratio 1.023 and radius ratio 1.639 for the components. These numbers clearly tells us that this binary is composed of two stars with different masses and the primary evolved so that its temperature is decreased while its radius is increased. However, the mass ratio of this star in \cite{Lee2015}'s list is given as $q=1.0$. Moreover, there are many binary star LCs with vague minima shapes due to scatter (i.e. ASAS 075422--1713.1) and their mass ratios in the \cite{Lee2015}'s work are given to be $q=1.0$. Because of these uncertainties, we have decided not to use \cite{Lee2015}'s results in this study.

From the perspective given in \S2, the candidate stars that could be twins in the ASAS catalog were identified in two steps. In the first step, all eclipsing binaries in the catalog were reviewed and systems with primary and secondary minimum depths being similar and with flat maximum light levels were chosen. The reason for this filtering is the determination of eclipsing binaries with components that have not evolved and did not fill their Roche lobes. If a system undergoes a mass transfer because of Roche lobe filling, initial mass ratio can not be maintained.

In the ASAS catalog, a total of 11038 eclipsing variable stars are categorized according to their variability types (e.g. EC, ESD, ED-types for eclipsing variables). Considering that categorization may be inaccurate for some light curves, the light curves of all eclipsing binary stars have been reviewed in the catalog without discrimination. In this way, in the first step, we obtained a list of candidates with 470 binary stars with similar minima depths and flat maximum. In the second step, a second list was created by removing candidates that could not be observed in $\sim$+37$^\circ$ N latitude, where our telescopes are located.

\begin{figure}
\includegraphics[width=\columnwidth]{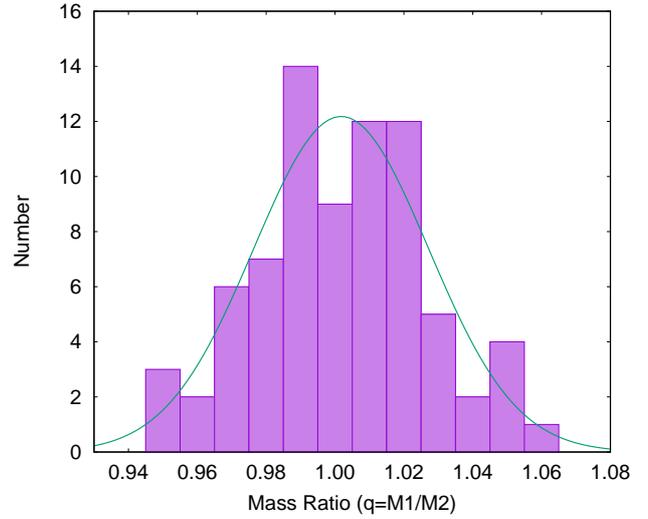}
\caption{Mass ratio distribution of twin binary candidates. According to the Gaussian fit, the a peak is at q($M_1/M_2$)=1.002$\pm$0.003.\label{fig:q_hist}}
\end{figure}

The light curve analysis of the identified 240 candidate stars was performed with {\sc phoebe} \citep{Prsa2005}. According to depth of minima and duration of eclipses of synthetic LCs, if they imply similar radii and temperatures, twin binaries were chosen. In the first stage of the analysis, we are interested only in the temperature ratio rather than the individual temperatures. Therefore, in case temperature of the components is not known, any temperature can serve as an input value. We adopted 8000K for the primary component for this purpose. The mass ratio was adopted as $q(M_1 / M_2) =$ 1. Both radii and temperatures were redetermined after obtaining spectral type information of the systems (see \S2.4 and \S2.5). Based on the obtained relative radius and temperature information of the components, the mass ratios were derived by using Eq.~\ref{eq:massratio}. It has been decided that the systems with mass ratio ($q$) between 0.94-1.06 could be selected as twins. In this selection, it is considered that \cite{LucyRicco1979} defined the ratio of the masses of the components for the twin stars as $q\cong$0.97, which was later updated by \cite{Tokovinin2000} to $q>$0.95. This range is also close to the approach of \cite{Pinsonneault2006}. With this mass ratio discrimination, the number of the twin candidates was reduced to 68. This is the final list which were planned for photometric and spectroscopic observations of this study (\S2.3).
In addition to this list of twins, there are 12 binary systems listed in Table~\ref{tab:catDEB_twins} which have photometric data in the ASAS catalogue and are also listed in the catalogue of detached main sequence binary systems \citep{Eker2018}. These latter systems are well studied in the literature and the physical parameters of the components are well known. The histogram of the mass ratios obtained in this study is shown in Fig.~\ref{fig:q_hist}.

\subsection{Observations and Data Reduction}

The twin candidates are observed at 4 different telescopes. The coordinates of the observatories and the names of the focal plane instruments are given in Table~\ref{tab:observatory}. Observations are divided into two groups as photometric and spectroscopic. Photometric observations were done in three different telescopes (T100, UBT60, ADYU60) and CCDs (SI1100, Alta U47, iKon-M 934) equipped with standard Johnson $UBVRI$ filters. All photometric observations, which have been performed in several observing runs between 2014-2017, were transformed to the standard system using the transformation equations given in Eq.~\ref{eq:transformation}. The transformation coefficients were found by analyzing the standard star observations, which are presented in Table~\ref{tab:transformation} together with the rms values of the linear fits. The standard stars were selected from the list of standards around celestial equator prepared by \cite{Landolt1992}.

\begin{table*}
\renewcommand{\thetable}{\arabic{table}}
\centering
\caption{Observatories and telescopes used.}
\begin{tabular}{ccccc}
\hline \hline
Observatory    & Location, Altitude  & Telescope  & Focal Plane Device & Number of Observed Objects \\
\hline
TUG & +36$^{\circ}$.82416 N, 30$^{\circ}$.33555 E, 2500m  & RTT150, 150cm & TFOSC & 54 \\
    &   & T100, 100cm & SI1100 & 60  \\
    &   & UBT60, 60cm  & Alta U47 & 25 \\
    &   & UBT60, 60cm  & eShel & 5 \\
ADYUO & +37$^{\circ}$.75192 N, 38$^{\circ}$.22542 E, 700m & ADYU60, 60cm & iKon-M 934 & 17\\
\hline \label{tab:observatory}
\end{tabular}
\end{table*}

\begin{equation}
\label{eq:transformation}
\begin{split}
(u-b) & =\phi(U-B)+\zeta_{ub}, \\
(b-v) & =\mu(B-V)+\zeta_{bv}, \\
(v-r) & =\lambda(V-R)+\zeta_{vr}, \\
(r-i) & =\beta(R-I)+\zeta_{ri}, \\
V-v & =\epsilon(B-V)+\zeta_v
\end{split}
\end{equation}


The uncertainties of standard magnitudes in many cases close to the uncertainty of the transformation model parameter are given in Table~\ref{tab:transformation}. This is because the magnitude reading by aperture photometry is dependent on S/N ratio by $\sigma=\frac{1.086}{S/N}$ which can be set to a certain value by arranging the exposure time. However, atmospheric effects have wavelength dependent variations on magnitudes and colors. Thus, the uncertainties given in Table~\ref{tab:transformation} represent the lower limit. In Table~\ref{tab:obs_data}, basic information together with observed photometric data of twins are presented.

\begin{table*}
\renewcommand{\thetable}{\arabic{table}}
\centering
\caption{Photometric transformation coefficients for each observing facility.}
\begin{tabular}{cccccc}
\hline \hline
& $\phi$  & $\mu$  & $\lambda$ & $\beta$ & $\epsilon$ \\
Telescope & $\zeta_{ub}$  & $\zeta_{bv}$  & $\zeta_{vr}$ & $\zeta_{ri}$ & $\zeta_{v}$ \\
& rms & rms & rms & rms & rms \\
\hline
UBT60 & 0.87$\pm$0.03 & 0.827$\pm$0.005 & 1.06 $\pm$ 0.01  &        1.14$\pm$0.01   & -0.079$\pm$0.009 \\
     & 2.52$\pm$0.02 & 0.022$\pm$0.003 & -0.051$\pm$0.005 &  -0.575$\pm$0.004 & -1.210$\pm$0.006\\
    &0.079 & 0.008 & 0.010 & 0.003 & 0.012 \\
T100  & 0.65$\pm$0.04 & 0.93$\pm$0.01 & 1.00$\pm$0.01 & 0.98$\pm$0.02 & 0.009$\pm$0.004 \\
       & 2.48$\pm$0.05 & 0.11$\pm$0.01 & 0.007$\pm$0.008 & -1.03$\pm$0.01  & 2.511$\pm$0.005 \\
        & 0.077 & 0.008 & 0.013 & 0.009 & 0.014 \\
ADYU60 & 0.95$\pm$0.07 & 0.734$\pm$0.008 & 0.904$\pm$0.017 & 1.244$\pm$0.020 & -0.054$\pm$0.010 \\
       & 3.05$\pm$0.07 & 0.166$\pm$0.008 & -0.105$\pm$0.012 & -0.890$\pm$0.017 & -5.01$\pm$0.01 \\
        & 0.185 & 0.021 & 0.031 & 0.041 & 0.026\\
\hline \label{tab:transformation}
\end{tabular}
\end{table*}

Spectroscopic observations were made by two \`{e}chelle spectrographs TUG Faint Object Spectrograph (TFOSC) and Shelyak Instruments eShel Spectrograph (SIeS) attached to the RTT150 and UBT60 telescopes, respectively. RTT150+TFOSC combination is capable of observing stars in \`{e}chelle mode down to 13th magnitude. It is attached to the telescope at the cassegrain focus. It provides spectrum between $\lambda\lambda$3900-9000 in 11 \`{e}chelle orders with a resolving power of R$\sim$5600. SIeS is fed by two fibers, one is for object and the other is for calibration which is used with a ThAr lamp. SIeS is equipped with a CCD camera that has 2184 x 1472 pixels of 6.8 x 6.8 $\mu$m each covering the wavelength range of $\lambda\lambda$4050-8160 with a resolving power of R $\sim$ 12000. UBT60+eShel combination is capable of observing stars down to 11th mag.

\begin{table*}
\caption{Basic information and photometric data for ASAS twins. The full table is available online.\label{tab:obs_data}}
\begin{tabular}{rccllrrrrrr}
\hline
No & ASAS & Cross & T$_0$ & P & V$_{obs}$ & $(U-B)$ & $(B-V)$ & $(V-R)$ & $(V-I)$ & D$_{GAIA}$ \\
   & ID & Reference & (HJD-2400000) & (d) & (mag) & (mag) & (mag) & (mag) & (mag) & (pc) \\
\hline
1 & 012424--2753.4 & CD-28 432 & 51870.2925 & 2.065400 & 10.54 & 0.37 & 0.46 & 0.26 & 0.26 & 506 \\
2 & 022311+1630.6 & TYC 1215-974-1 & 52635.6689 & 0.909741 & 11.49 & 0.72 & 0.85 & 0.57 & 0.55 & 129 \\
3 & 023039--1253.9 & BD-13 466 & 51884.9172 & 5.757590 & 10.51 & 0.00 & 0.58 & 0.32 & 0.28 & 271 \\
4 & 024458--1746.8 & FK Eri & 51869.2225 & 2.232229 & 9.10 & 0.05 & 0.46 & 0.30 & 0.27 & 260 \\
5 & 024644+0107.9 & HS Cet & 51919.6024 & 3.690981 & 10.43 & 0.04 & 0.43 & 0.23 & 0.25 & 376 \\
\hline
\end{tabular}
\end{table*}

\subsection{Modeling the Observed Spectra}

\begin{figure}
\begin{center}
\begin{tabular}{c}
\includegraphics[width=\columnwidth]{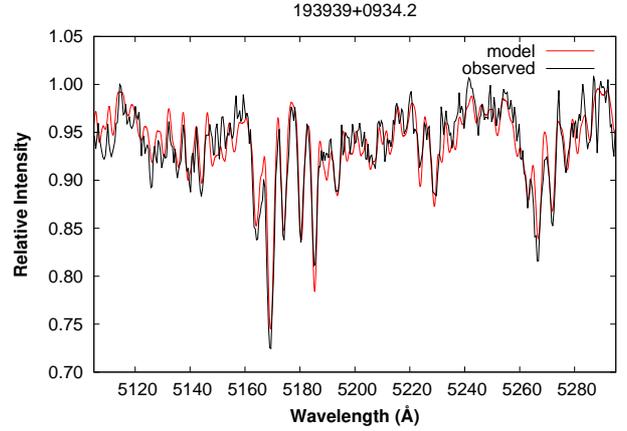}
\end{tabular}
\end{center}
\caption{Model atmosphere fitted to the observed composite spectrum of 193939+0934.2 at orbital phase $\phi=$0.74.\label{fig:atm_models}}
\end{figure}

If a spectrum exists for a twin, some physical parameters related to its atmosphere such as surface temperature, gravity, metallicity, projected rotation velocity, micro and macro-turbulence velocity could be obtained by modeling the spectral features. In this study, the spectrum of the target stars were obtained at orbital phases close to quadrature ($\phi\sim$0.25 or 0.75), which allowed us to see and model spectral features of both components as well as their relative velocities ($\Delta V=|V_1 - V_2|$). Temperature of the targets in our list are less than 15000K, so \cite{Kurucz1993} atmosphere models and related codes ({\sc atlas9, synthe}) which are based on LTE approximation were adopted for modeling stellar spectra. Spectral lines were modeled in the following steps: a- Models were calculated for certain range values of atmospheric parameters (constructing a grid of synthetic spectra), b- models were shifted along the wavelength axis, taking into account the orbital phases and light contribution ratios obtained from the light curve analysis (constructing the composite spectra), c - the compatibility of the synthetic composite spectrum with the observed spectrum was examined (comparing models with observations), this cycle was continued until the best fit was achieved. In Fig.\ref{fig:atm_models}, atmosphere models for a twin (ASAS ID 203251-1841.6) in our list is shown. The atmospheric parameters derived for each twin are given in Table~\ref{tab:twin_parameters}.

\subsection{Interstellar Extinction Towards The Observed Twins}

The observed colors of twins allowed us to locate their location in the color-color (CC) diagrams. Thus, we were able to find the color excesses of a related twin binary using the galactic extinction law in the form of a ratio of color excesses (e.g. $\frac{E(U-B)}{E(B-V)}=0.72$). The galactic interstellar reddening law have been adopted from \cite{Neckel1980} to be 0.72, 1.65 and 0.82 for the CC ratios, $E(U-B)/E(B-V)$, $E(B-V)/E(V-I)$ and $E(B-V)/E(V-R)$, respectively. Once the color excesses have been found, the true colors of the systems could have been derived and therefore spectral types could be determined. The spectral types determined are given in Table~\ref{tab:twin_parameters}.

The systems with atmospheric parameters are spectroscopically observed systems and their spectral types were determined directly from their spectra. For other systems, spectral types are determined by matching the unreddened colors with the $UBV$ data for spectral classes O5 to M8 given by \cite{Fitzherald1970}. 

Rarely, we run into the cases that spectral type found from one of the CC diagrams does not match with the results found from the other CC diagrams. This is due to the varying uncertainty of the observed colors. Especially, the uncertainty in the $U-B$ color is on the order of 0.15 mag in average (see Table~\ref{tab:transformation}) which changes nightly. In that case, the spectral type is determined from the other CC diagrams with sufficient certainty. Thus, a quality remark was found to be necessary to grade the spectral type data according to the method being used. If spectroscopy is available and spectral type is determined by spectral modeling the grade is A. Depending on the consistency of models with spectral data grade A is divided into A+ and A-. If spectral type is determined by three CC diagrams, the grade is B, if spectral type is supported by two CC diagrams, the grade is C, and if spectral type is supported by only one CC diagram, the grade is D.

In some cases, due to the uncertainty of the observed color, the star's extinction line does not intersect the curve of the unreddened main sequence stars. If this is the case for three CC diagrams and there is no spectroscopic observations of that star, then the spectral type is adopted from the observed color corresponding the spectral type given by the GAIA (\cite{Gaia2016}, \cite{Gaia2018}) input catalogue. In this case the grade of the data is given as E.

\begin{table*}
\caption{Results from the analysis of photometric and spectral data. Orbital phase is calculated using the ephemeris given in Table~\ref{tab:obs_data}. The full table is available online.}\label{tab:twin_parameters}
\begin{tabular}{rccccrrrcccc}
\hline
No & ASAS & $T_{eff1}$ & $T_{eff2}$ & $log\,g$ & $r_1/r_2$ & $M_1/M_2$ & V$_{rot}$ & Phase & $\Delta$V &  Sp.  & Q \\
 & ID & (K) & (K) & (cgs) &  &  & (km s$^{-1}$) & $\phi$ & (km s$^{-1}$) &  & \\
\hline
1 & 012424--2753.4 & 6397 & 6362 & & 0.977 & 0.994 & & & & F5V & D \\
2 & 022311+1630.6 & 4534 & 4527 & 4.49 & 1.076 & 1.027 & 50 & 0.158 & 244 & K4V - G5V & A$^-$ \\
3 & 023039--1253.9 & 6792 & 6928 & 4.27 & 1.014 & 0.988 & 50 & 0.801 & 126 & F5V & A$^-$ \\
4 & 024458--1746.8 & 6792 & 6807 & 4.28 & 1.000 & 0.998 & 50 & 0.184 & 208 & F2V & A \\
5 & 024644+0107.9 & 6310 & 6298 & 4.27 & 0.964 & 0.985 & 50 & 0.713 & 181 & F4V & A \\
\hline
\end{tabular}
\end{table*}

\begin{table*}
\caption{Twins in the list of absolute parameters of 509 main-sequence stars  \citep{Eker2018}.}\label{tab:catDEB_twins}
\begin{tabular}{rccccrrc}
\hline
No & ASAS & $T_{eff1}$ & $T_{eff2}$ & $log\,g_{1,2}$ & $r_1/r_2$ & $M_1/M_2$ &  Sp.  \\
 & ID & (K) & (K) & (cgs) &  &  &  \\
\hline
1 & 022627+1253.9 & 10300 & 9800 & 4.075, 4.137 & 1.076 & 0.993 & A0V \\
2 & 030807--2445.6 & 4100 & 4055 & 5.000, 4.632 & 1.064 & 0.981 & K7V \\
3 & 045304--0700.4 & 5340 & 5125 & 4.503, 4.515 & 1.018 & 0.993 & K2/4V \\
4 & 052821+0338.5 & 5103 & 4751 & 4.050, 4.080 & 1.058 & 0.972 & K1/3V \\
5 & 082552--1622.8 & 4350 & 4090 & 4.600, 4.580 & 0.993 & 0.978 & K \\
6 & 084108--3212.1 & 7468 & 7521 & 4.010, 4.126 & 1.163 & 0.965 & A8/9V \\
7 & 093814--0104.4 & 4360 & 4360 & 4.550, 4.554 & 1.004 & 1.004 & K \\
8 & 095707--0120.7 & 6440 & 6160 & ... & 0.993 & 1.024 & F3/5V \\
9 & 103821+1416.1 & 6129 & 5741 & 4.54, 4.31 & 0.800 & 0.959 & G0/2V \\
10 & 111245+0020.9 & 6316 & 6190 & 4.124, 4.189 & 1.091 & 0.976 & F9V \\
11 & 135035--5830.0 & 8000 & 8280 & 3.487, 3.678 & 1.257 & 0.983 & A7V \\
12 & 214138+1439.5 & 7850 & 7600 & 4.010, 4.100 & 1.128 & 0.983 & A7/8V \\
\hline
\end{tabular}
\end{table*}

\section{A case study: An A-type twin binary HD154010}

Among the twins listed in Table~\ref{tab:twin_parameters}, HD154010 (ASAS 170158+2348) is one of the systems selected for systematic observations. As far as we are aware like most of the twins in our list, there are no studies related to HD~254010 in the literature. It is even not listed as a variable in the {\sc simbad} database\footnote{http://simbad.u-strasbg.fr/simbad/}.

Spectra of HD154010 have been obtained in 8 nights between August and October of 2018 by using SIeS that is attached to UBT60 telescope. Each night, six spectra in series, each with exposure times of 600s, have been collected and they have been combined in order to obtain sufficient ($\sim100$) S/N ratio. A standard procedure has been applied for the reduction of the spectra which included bias and dark extraction, flat field normalization, wavelength calibration and continuum correction. The image reduction and analysis has been performed in the {\sc iraf} \citep{Tody1993} platform.

\begin{figure}
\includegraphics[width=\columnwidth]{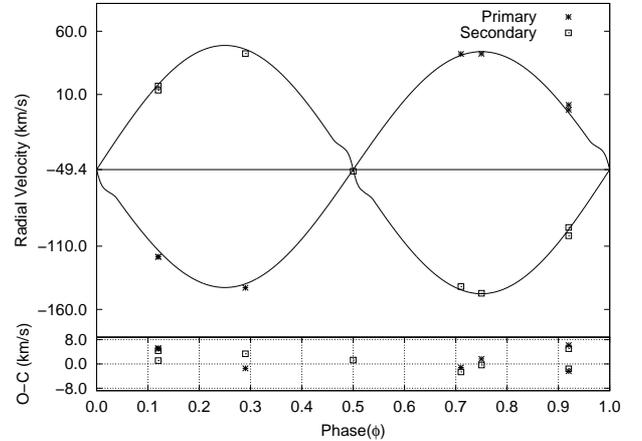}
\caption{Spectroscopic orbital solution of HD~154010. The residuals from the fit are shown in the bottom panel.\label{fig:rv}}
\end{figure}

In \S2, the system is identified as a binary composed of two slowly rotating late A spectral type stars (see Table~\ref{tab:twin_parameters}). This spectral type stars are known to show a lot of metallic lines which is good for reliable RV measurements. The measuring individual RVs have been performed by fitting Gaussian to the central part of the individual spectral lines. After several fittings on different spectral lines, a final RV is adopted by averaging the measurements and applying the  Heliocentric RV correction. Final RVs are listed in Table~\ref{tab:rv} and the best fitting spectroscopic orbital model parameters are given in Table~\ref{tab:rvpar}. In Fig.~\ref{fig:rv}, the orbital model is shown.

\begin{table}
\renewcommand{\thetable}{\arabic{table}}
\centering
\caption{The RVs of the components of HD154010.}
\begin{tabular}{ccrr}
\hline \hline
HJD      & Phase  & \multicolumn{1}{c}{RV$_{1}$}  & \multicolumn{1}{c}{RV$_{2}$} \\
-2400000 & \multicolumn{1}{c}{$\phi$} & (kms$^{-1}$) & (kms$^{-1}$) \\
\hline
58356.34047 & 0.92 & -2.1   & -95.3  \\
58357.28471 & 0.12 & -118.3 & 16.7   \\
58396.23763 & 0.29 & -142.9 & 42.3   \\
58397.23345 & 0.50 & --     & -50.7  \\
58398.24159 & 0.71 & 42.1   & -141.9 \\
58399.22781 & 0.92 & 1.7    & -101.8 \\
58400.22453 & 0.12 & -118.5 & 13.4   \\
58403.22682 & 0.75 & 42.5   & -147.2 \\
\hline \label{tab:rv}
\end{tabular}
\end{table}

\begin{table}
\small
\begin{center}
\caption{Parameters of the spectroscopic orbit.}
\begin{tabular}{lc}  \hline \hline
Parameter                   & Values       \\
\hline
$P$(days)                   &   4.76988  \\
$T_{0}$(HJD-2458353)        & 0.1359 $\pm$ 0.0086 \\
$K_{1}$(km s$^{-1}$)        & 93.47 $\pm$ 0.10  \\
$K_{2}$(km s$^{-1}$)        & 98.39 $\pm$ 0.10  \\
$e$                         & 0.0 (fixed) \\
$V_{\gamma}$(km s$^{-1}$)   &  -49.41 $\pm$ 0.70  \\
$q (M_2/M_1)$               &   0.948 $\pm$ 0.023 \\
$m_{1}{\mathrm sin}^{3}i (M_{\odot})$ &  1.76 $\pm$ 0.02 \\
$m_{2}{\mathrm sin}^{3}i (M_{\odot})$ &  1.67 $\pm$ 0.02 \\
$a {\mathrm sin}i (R_{\odot})$     & 17.88 $\pm$ 0.12 \\
\hline \label{tab:rvpar}
\end{tabular}
\end{center}
\end{table}

\begin{figure*}
\begin{center}
\begin{tabular}{cc}
\includegraphics[width=\columnwidth]{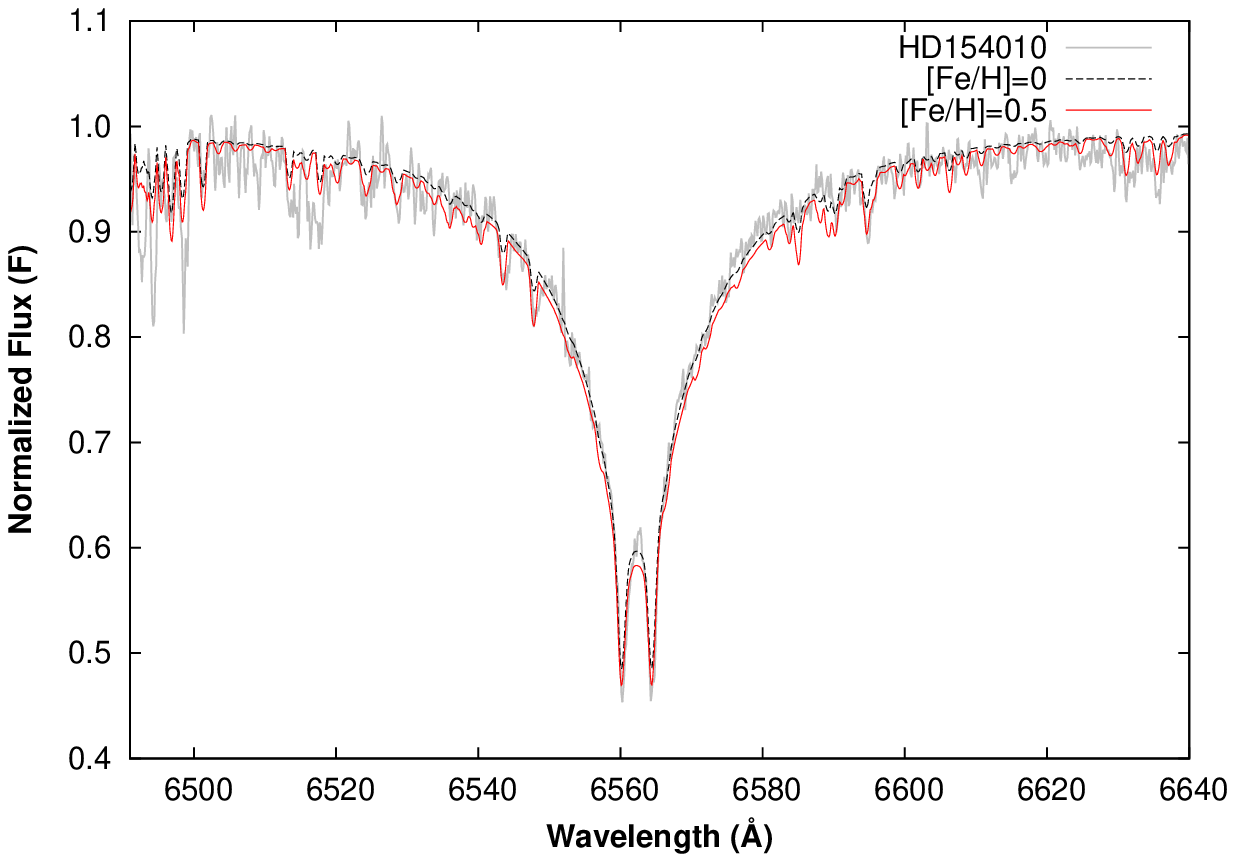} &
\includegraphics[width=\columnwidth]{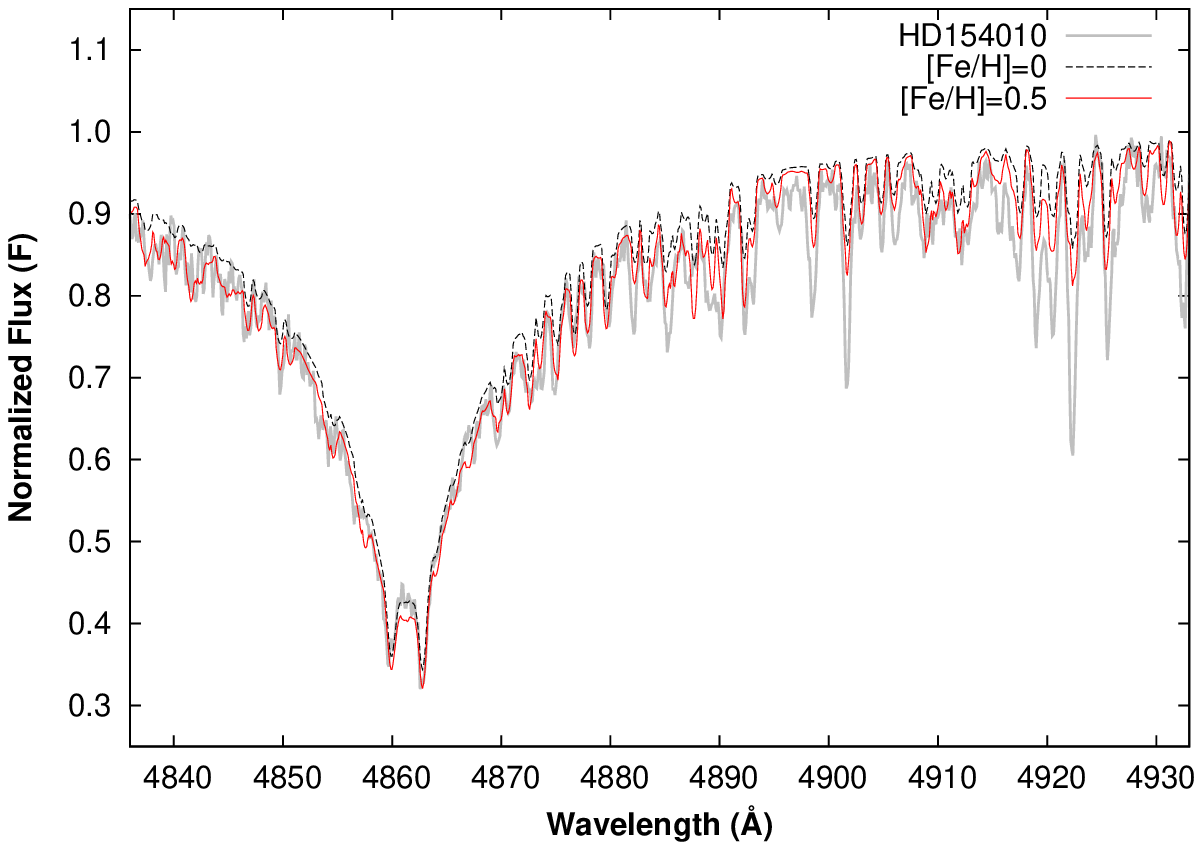}
\end{tabular}
\end{center}
\caption{Atmosphere models fitted to H$_\alpha$ and H$_\beta$ lines in the composite spectrum of HD~154010.\label{fig:balmers}}
\end{figure*}

\begin{figure}
\includegraphics[width=60mm, angle=270]{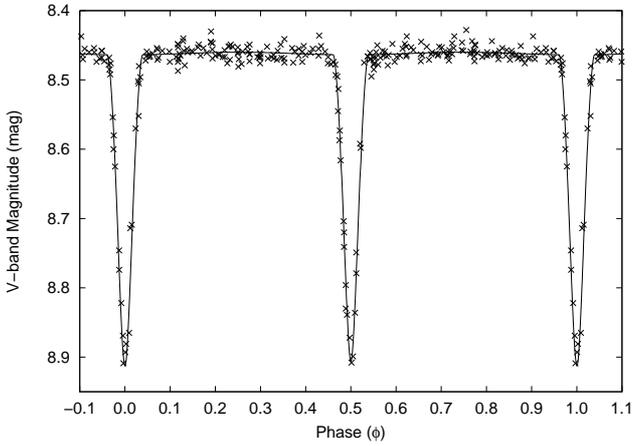}
\caption{Light curve model fitted to V-band data of HD~154010.\label{fig:lc}}
\end{figure}

Modeling the observed spectra around $H_\alpha$ and $H_\beta$ lines (Fig.~\ref{fig:balmers}) allowed us to determine the temperature of the components, which are later improved with light curve analysis (Fig.~\ref{fig:lc}). Combining the LC model parameters (Table~\ref{tab:lcpar}) with spectroscopic model parameters (Table~\ref{tab:rvpar}) enabled us to derive the absolute physical parameters of the system, which are given in Table~\ref{tab:absolutepar}. Both LC and spectral analysis showed that there is no third light effect in the system. The mass of the components of HD~154010 imply spectral type of A5 in the main-sequence, however, temperature, radii and surface gravity of the components show that the system is not a main-sequence but a sub-giant luminosity class with A7 spectral type.

\begin{table}
\begin{center}
\caption{V-band LC model parameters of HD154010.} \label{tab:lcpar}
\begin{tabular}{lcc}\hline\hline
Parameter              &  Value & Error\\
\hline
$P$ (days)             &\multicolumn{2}{c}{4.76988}  \\
$T_0$ (HJD-2452709)    & 0.1630 & 0.0011 \\
$T_{\rm eff1}(K)$      &  \multicolumn{2}{c}{7600}\\
$T_{\rm eff2}(K)$      &   7621 & 35   \\
$q$                    &  \multicolumn{2}{c}{0.95} \\
$L_{1}/L_{1+2}$ ($V$)  &   0.50 & 0.02 \\
$e$                    &  \multicolumn{2}{c}{0.0}  \\
$w$ $(^{o})$           & \multicolumn{2}{c}{0.0}  \\
$i$ $(^{o})$           &   85.8 & 0.1   \\
$\Omega_{\rm 1}$       &   8.613 & 0.224  \\
$\Omega_{\rm 2}$       &   8.837 & 0.158  \\
${r}_{\rm 1}$      &   0.131 & 0.004 \\
${r}_{\rm 2}$      &   0.122 & 0.004 \\
$\chi^{2}$         &  \multicolumn{2}{c}{0.005} \\
\hline
\end{tabular}
\end{center}
\end{table}

\begin{table*}
\small
\caption{Binary stellar parameters of HD 154010. Errors of parameters are given in parenthesis.} \label{tab:absolutepar}
\begin{tabular}{lccc}\hline
Parameter                          & Symbol  & Primary & Secondary\\
\hline
Spectral type                      & Sp      &  A8 V   &  A8 V \\
Mass (M$_\odot$)                   & \emph{M}& 1.80(0.01) & 1.71(0.01)\\
Radius (R$_\odot$)                 & \emph{R}& 2.37(0.08) & 2.21(0.04)\\
Separation (R$_\odot$)             & \emph{a}& \multicolumn{2}{c}{18.13(0.02)} \\
Orbital period (days)              & \emph{P}       &
\multicolumn{2}{c}{4.76988}   \\
Orbital inclination ($^{\circ}$)   & \emph{i}       & \multicolumn{2}{c}{85.82(0.02)}  
\\
Mass ratio                         & \emph{q}       &
\multicolumn{2}{c}{0.95(0.02)}\\
Eccentricity                       & \emph{e}       & \multicolumn{2}{c}{0.0
	(fixed)}  \\
Surface gravity (cgs)              & $\log g$       & 3.945(0.031)& 3.984(0.019)  
\\
Combined visual magnitude (mag)  & \emph{V}       & \multicolumn{2}{c}{8.46}\\
Individual visual magnitudes (mag) & \emph{V$_{1,2}$}       & 9.12(0.05) &
9.13(0.05)  \\
Combined colour index (mag)      & $B-V$         &  \multicolumn{2}{c}{0.28(0.02)}
\\
Temperature (K)                    & $T_{\rm eff}$ & 7300(100)    & 7300(100)         \\
Luminosity (L$_\odot$)             & $\log$ \emph{L}& 1.158(0.050) &  1.096(0.043)\\
Bolometric magnitude (mag)         &$M_{\rm bol}$& 1.855(0.125) & 2.010(0.108)  \\
Absolute visual magnitude (mag)    &$M_{\rm v}$  & 1.820(0.125) & 1.975(0.108)  \\
Bolometric correction (mag)        &\emph{BC}& 0.035(0.001)     & 0.035(0.001) \\
Velocity amplitudes (km\,s$^{-1}$)  &$K_{\rm 1,2}$& 93.47(0.10) & 98.39(0.10) \\
Systemic velocity (km\,s$^{-1}$)    &$V_{\gamma}$ & \multicolumn{2}{c}{-49.5(0.7)} \\
Computed synchronization velocities (km\,s$^{-1}$)& V$_{\rm synch}$ & 25.1(0.8) &
23.4(0.4) \\
Observed rotational velocities (km\,s$^{-1}$) & V$_{\rm rot}$ & 25(5) & 25(5)\\
Distance (pc) & \emph{d} & \multicolumn{2}{c}{279(22)}\\
\hline
\end{tabular}
\end{table*}

\section{Discussion and Conclusion}

Using the spectral types given in Table~\ref{tab:twin_parameters}, we know the spectral type distribution of the twins. It is useful to make an inter-comparison of ASAS twins with the detached binaries in the same catalogue. Thus, selection effects will cancel out and we will see a real comparison of twins with the detached binaries. Such a comparison is expected to give an answer to the question if twins have different formation mechanisms than detached binary systems.

For this purpose, binaries that were classified as detached in the ASAS catalogue have been sorted. The number of such binaries determined by the ASAS team with the "ED" code is 2275. The temperature of these binaries can be found in the GAIA (\cite{Gaia2016}, \cite{Gaia2018}) catalogue as discussed in \S2.5. However, the color-temperature calibration of the GAIA mission is limited to stars with spectral types later than A0. Including this limitation, missing temperatures in the GAIA database reduces the number of detached binaries with known spectral types to 1448. Unfortunately, the number of early type detached binary stars in the ASAS database is not known.

\begin{figure}
\includegraphics[width=\columnwidth]{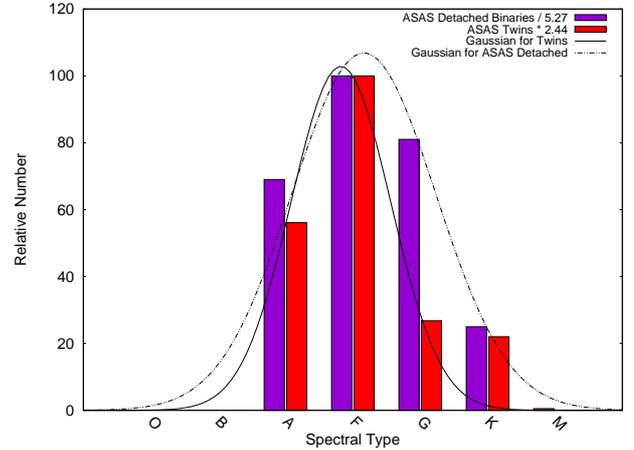}
\caption{The normalised spectral distribution of detached binaries together with twin binaries in the ASAS catalogue. For each distribution the Gaussian fit is shown.\label{fig:sp_hist}}
\end{figure}

Using this information and the spectral types given in Table~\ref{tab:twin_parameters}, histogram distribution of the twin binaries is generated in Fig.~\ref{fig:sp_hist}. In the figure, all data are normalized for a better comparison. The normalization constants are given in the same figure. As can be seen from the Fig.~\ref{fig:sp_hist}, the maximum concentration of both distribution indicates F-spectral type with twins of narrower Gaussian.

We can interpret from Fig.~\ref{fig:sp_hist} that formation mechanism of twin binaries with respect to spectral type or mass is similar to the formation mechanism of the detached binaries.  The missing part (OB stars) in the ASAS detached binary list caused a shift in the maximum of the distribution to later spectral types. The number of known OB twins are less than the number of G-K-M twins. Therefore, we do not think the existance of OB twins in our Fig.~\ref{fig:sp_hist} will change the general gaussian distribution of twins substantially. Nevertheless, the existance of OB detached binaries will shift the gaussian towards eary spectral type region which will lead to approach of both distributions. Apparently \cite{Bate1998}'s description is in play; secondary accumulates more material as it moves on a larger orbit, so detached binary formation prefers twin binaries, making their mass ratio closer to one. However, this does not exclude the mechanism of \cite{Tokovinin2000} that twin binaries occurring in multiple systems. Thus, we encourage investigations of additional components and kinematical search of their origin which may end up in OB-associations or open clusters where multiplicity occurs frequently.


The fact that the ASAS catalog does not encounter twins of the early spectral type does not mean that they do not exist. Recent studies (e.g. \cite{Bakis2007}, \cite{Bakis2013}) and the \cite{Eker2018} point out that their number may be as many as the late spectral type twin binary stars.

\section*{Acknowledgements}
This work has been supported by The Scientific and Technological Research Council of Turkey (T\"{U}B{\.{I}}TAK) under the project code 115F029. We thank to T\"{U}B{\.{I}}TAK for a partial support in using T100 telescope with project number 14CT100-692. This research has made use of the SIMBAD database, operated at CDS, Strasbourg, France. This work has made use of data from the European Space Agency (ESA) mission {\it Gaia} (\url{https://www.cosmos.esa.int/gaia}), processed by the {\it Gaia} Data Processing and Analysis Consortium (DPAC,
\url{https://www.cosmos.esa.int/web/gaia/dpac/consortium}). Funding for the DPAC has been provided by national institutions, in particular the institutions participating in the {\it Gaia} Multilateral Agreement.




\bibliographystyle{mnras}
\bibliography{references} 



\bsp	
\label{lastpage}
\end{document}